\documentclass[epj]{svjour}
\usepackage{graphics}
\usepackage{color}
\usepackage{bm}
\usepackage{ulem}

\newcommand{\sss}{\mathop{\bf A}\limits}
\newcommand{\bsigma}{\boldsymbol{\sigma}}
\newcommand{\modifym}[2]{#2}
\begin{document}
\title{Coupling Finite Element Method with Large Scale Atomic/Molecular Massively Parallel Simulator (LAMMPS)\\
for Hierarchical Multiscale Simulations}
\subtitle{Modeling and simulation of amorphous polymeric materials}
\author{Takahiro Murashima\inst{1}, Shingo Urata\inst{2} \and Shaofan Li\inst{3}
}                     
%
%
\institute{Tohoku University, \email{murasima@cmpt.phys.tohoku.ac.jp} \and AGC Inc.,
\email{shingo-urata@agc.com
} \and University of California Berkeley, \email{shaofan@berkeley.edu}}
\date{Received: date / Revised version: date}
%
\abstract{
In this work, we have developed a multiscale computational algorithm
to couple finite element method with an open source molecular dynamics code --- the
Large scale Atomic/Molecular Massively Parallel Simulator (LAMMPS) ---
to perform hierarchical multiscale simulations in highly scalable parallel computations.
The algorithm was firstly verified by performing simulations of single crystal copper deformation,
and a good agreement with the well-established method was confirmed.
Then, we applied the multiscale method to simulate mechanical responses
of a polymeric material composed of multi-million fine scale atoms inside the representative
unit cells (r-cell) against uniaxial loading.
It was observed that the method can successfully capture plastic deformation in the polymer at macroscale,
and reproduces
\modifym{}{the double yield points typical in polymeric materials,}
strain localization and necking deformation after \modifym{the elastic limit}{the second yield point}.
In addition, parallel scalability of the multiscale algorithm was examined
up to around 100 thousand processors with 10 million particles,
and an almost ideal strong scaling was achieved thanks to LAMMPS parallel architecture.
} 

\authorrunning{T. Murashima, S. Urata, and S. Li}
\titlerunning{Hierarchical multiscale coupling algorithm between FEM and LAMMPS}

\maketitle
\section{Introduction}
\label{intro}
Multiscale simulation methods, which couple atomistic models with continuum models,
are advanced computational technologies aimed to understand material properties
under specific conditions without using any empirical or experimental information.
\modifym{}{
The multiscale approach is not simply a computational convenience, but deeply rooted
in material physics,}  
because almost all physical events in nature are inherently multiscale phenomena
with different time and length scales.
In past two decades, a variety of multiscale techniques have been proposed
and developed e.g. \cite{tadmor2011,liu2004,To2005,Murashima2013}.
In general, the multiscale methods in condensed matters may be categorized 
into two major classes;
the concurrent multiscale modeling and the hierarchical multiscale modeling.

The \modifym{con-current}{concurrent} multiscale approach directly connects continuum 
and atomistic models by communicating
displacement or force of atoms to nodes, particles or quadrature points of continuum body
on-the-fly to link microscale and macroscale.
Since the latter half of the 1990s, several pioneering multiscale methods 
on coupling atomistic methods with finite elements
have extensively studied.
Broughton et al. have conducted fracture simulation 
of Silicon by coupling finite element method (FEM),
molecular dynamics (MD) simulation, and tight binding quantum calculation though
a handshake Hamiltonian \cite{abraham1998,broughton1999}.
The bridging scale method uses a projection operator to decompose 
the total displacement field into
mutually orthogonal coarse and fine scales, 
leading to a coupled multiscale equations of motion
in the MD and FEM models respectively, e.g. \cite{wagner2003,xiao2004}.
Recently, Li and his co-workers have proposed a novel decomposition 
method of MD simulation,
namely multiscale micromorphic molecular dynamics (MMMD), to model transition zone
between microscale and macroscale thorough mesoscale region \cite{li2016}.
The method allows us to couple MD simulations and FEM \cite{urata2016} 
or Peridynamics \cite{tong2016},
which usually models nonlocal continuum media by using particle discretization.

The other category of multiscale modelings is hierarchical coupling method,
which locally embeds fine scale models into coarse scale models.
A conventional procedure to build a hierarchical modeling
is adopting micromechanics homogenization approach by
employing a so-called representative volume elements (RVE) 
e.g. \cite{sheng2004,valavala2009,wernik2014}.
One of the most successful hierarchical
multiscale approaches is the quasicontinuum (QC) method proposed by Tadmore et al.,
in which the finite element mesh covers the entire simulation system,
and the mesh may be scaled-down to atomic dimensions at selected locations,
in which the fine scale modeling is in terms of molecular statics or energy minimization
or optimization \cite{tadmor1996,tadmor1999,miller2002}.
\modifym{}{
A similar but much simpler hierarchical approach is the
so-called Cauchy-Born rule (CBR) method, which assumes affine deformation 
in crystal lattice,
and it thus allows us to couple atomistic and continuum models directly through
constant deformation gradient and direct bottom-up upscale.
}
Thus, CBR was in principle only 
applicable to crystalline materials,
because it is based on the assumption of uniform deformation of crystal lattices; 
however, the idea has been further extended to nonuniform deformation in crystalline solids
by considering higher order deformation gradients e.g. \cite{sunyk2003,weinan2007}.
For instance, the higher order CBR has been combined
with crystal defect dynamics \cite{lyu2017,lyu2019} and
cohesive zone model \cite{urata2017a} to investigate
more complicated dislocation pattern dynamics and fracture mechanics of crystals.

Recently, the present authors have extended the idea of the lattice statics-based
Cauchy-Born rule to amorphous solids to conduct hierarchical multiscale modeling
of inelastic deformation in an amorphous solid \cite{urata2018}.
The method was called as the coarse-grained (CG) Parrinello-Rahman (PR) method,
since the molecular statics in the representative unit cell of the fine scale model
is used analogous to that of PR-MD simulation \cite{parrinello1980}.
The CG-PR method was systematically validated by comparing its numerical simulation results
with that of MD simulations,
and the CG-PR method was applied to study
the mechanical responses of a Lennard-Jones (LJ) 
binary glass \modifym{model}{and amorphous silica models} \cite{urata2017b}.
It has \modifym{be}{been} shown that the CG-PR method can successfully
reproduce shear band formation in
a single-notched amorphous solids using relatively coarse mesh
FEM models at macroscale \cite{urata2018}.
The advantage of CG-PR method is its independence of the constitutive empiricism, or
in other words, it does not need any ad hoc empirical modelings of complicated
constitutive relation of amorphous solids.
However, a critical drawback of the method is its expensive computation cost,
because a representative unit cell (r-cell) composed of many atoms
is needed for each and every quadrature points of a FEM model,
\modifym{}{
thus it requires running an enormous number of atomistic simulations all together at 
the same time.
Thus,
it calls an effective parallel computation algorithm for 
the proposed multiscale method,
which is crucial for any practical application
of such method.}

In this work, to improve computational efficiency of our in-house CG-PR code,
the CG-PR algorithm was coupled with the large scale parallel MD code, namely the
large scale atomic/molecular massively parallel simulator (LAMMPS) \cite{plimpton1995,LAMMPS}.
To do so, we first develop a coupling algorithm for a single MD cell,
whose shape is represented by an upper triangular stretch matrix of six components,
instead of general non-symmetric r-cell mentioned in section \ref{sec:1}.
In section \ref{sec:2}, we first
discuss the validation of the multiscale algorithm and multiscale code,
which were first validated by performing
a numerical simulation of deformation of crystalline copper.
Then, we conducted a large-scale parallel computation 
for simulation of inelastic deformation
of a polymeric material
to demonstrate applicability of the method, which is implemented in
a massively parallel supercomputer.
Finally, in Section \ref{sec:3}, we summarize and conclude the work.

\section{Computational Methods}
\label{sec:1}

We first discuss coupling between molecular dynamics (MD) with
finite element method (FEM).

\subsection{Coupling MD and FEM}
\label{sec:1.1}

To start, we first prepare a representative unit
cell (r-cell) whose shape is represented by a second order tensor or a matrix ${\bf H}$.
Atom position in the r-cell can be represented by a scaled atom position vector
(${\bf S}_i(t)$) for atom $i$,
and then the current position of the atom $i$ at time $t$ can be written as,
\begin{equation}
{\bf r}_i(t) = {\bf H}(t) \cdot  {\bf S}_i(t).
\label{eq:LAMMPS1}
\end{equation}
Employing CBR, we couple the deformation of
an r-cell within an element in a continuum model,
we may express the cell shape tensor as,
\begin{equation}
{\bf H}_e(t) = {\bf F}_e(t) \cdot {\bf H}_e(0),
\end{equation}
where ${\bf F}_e$ is deformation gradient tensor of an element $e$,
and ${\bf H}_e(0)$ is an initial shape of the r-cell.
Thus, the current position of an atom in an r-cell of the given $e$-th element is
expressed as,
\begin{equation}
{\bf r}_i^e(t) ={\bf H}_e(t) \cdot {\bf S}_i(0) = {\bf F}_e(t) \cdot {\bf H}_e(0) \cdot  {\bf S}_i(0).
\end{equation}

To apply a general MD code for hierarchical multiscale modeling,
the shape tensor ${\bf H}_e(t) $ should correspond to shape of an MD unit cell.
However, the shape tensor ${\bf H}_e(t) $ is not symmetric and
composed of independent nine components,
while MD unit cell in LAMMPS should be upper triangular matrix with six components.
\modifym{}{The shape tensor can be decomposed to a stretch tensor and a rotation tensor.
The stretch tensor, represented by the upper triangular matrix, satisfies the periodic
boundary conditions of the initial orthogonal basis, whereas the rotation tensor violates
the periodic boundary conditions.
We need to treat the stretch tensor and the rotation tensor separately
to implement the general purpose MD code to the multiscale simulation in the following way.
}

In our algorithm, we therefore decompose the deformation gradient tensor ${\bf F}_e$
into an orthogonal rotation matrix ${\bf Q}_e$ and an upper triangular stretch matrix ${\bf R}_e$ as follows.
\begin{equation}
{\bf F}_e(t) ={\bf Q}_e(t) \cdot {\bf R}_e(t).
\end{equation}
This process is conventional QR decomposition,
which is calculated by the Gram-Schmidt orthogonalization.
Note that the triangular stretch matrix ${\bf R}$ should be positive definite.
Otherwise, the stretch matrix ${\bf R}$ will introduce wrong axial inversion.

Instead of directly using ${\bf F}_e(t)$, the upper triangular stretch matrix ${\bf R}_e$
can be used as a deformation matrix of the MD unit cell as,
\begin{equation}
{\bf H}_{\rm MD}(t) = {\bf R}_e(t) \cdot {\bf H}_e(0),
\end{equation}
where ${\bf H}_{\rm MD}(t)$ represents shape of the MD unit cell at time $t$.
It is therefore, in a MD code, atomic coordinate should be
\begin{equation}
{\bf r}_i^{\rm MD}(t) = {\bf R}_e(t) \cdot {\bf H}_e(0) \cdot {\bf S}_i(0).
\end{equation}

Once we know the current atom coordinate, ${\bf r}_i^{\rm MD}(t)$,
a MD code can calculate the stress of the MD unit cell ($\bsigma_{\rm MD}(t)$)
using an interatomic interaction potential.
The stress in MD coordinate system is transferred to the real coordinate system of FEM
by using the rotation matrix ${\bf Q}_e$ as follows,
\begin{equation}
\bsigma_{\rm FEM}(t) = {\bf Q}_e(t) \cdot \bsigma_{\rm MD}(t)  \cdot {\bf Q}_e^T(t).
\end{equation}
In FEM code, the first Piola-Kirchhoff stress is used to evaluate a force acting on each node.
\begin{equation}
{\bf P}_{\rm FEM}(t) = J \bsigma_{\rm FEM}(t)  \cdot {\bf F}_e^{-T}(t),
\label{eq:Pmd}
\end{equation}
where $J$ is the Jacobian of the deformation gradient tensor ${\bf F}$.

In our multiscale code, we used LAMMPS as the MD calculation engine.
Because LAMMPS memorizes previous deformation matrix, ${\bf R}(t-1)$,
our FEM code provides the following differentiation between current
and previous upper triangular stretch matrices to LAMMPS.
\begin{equation}
\Delta{\bf R}_e(t) = {\bf R}_e(t)  \cdot {\bf R}_e^{-1}(t-1).
\label{eq:LAMMPS2}
\end{equation}

\begin{figure*}
\begin{center}
\resizebox{0.65\textwidth}{!}{
  \includegraphics{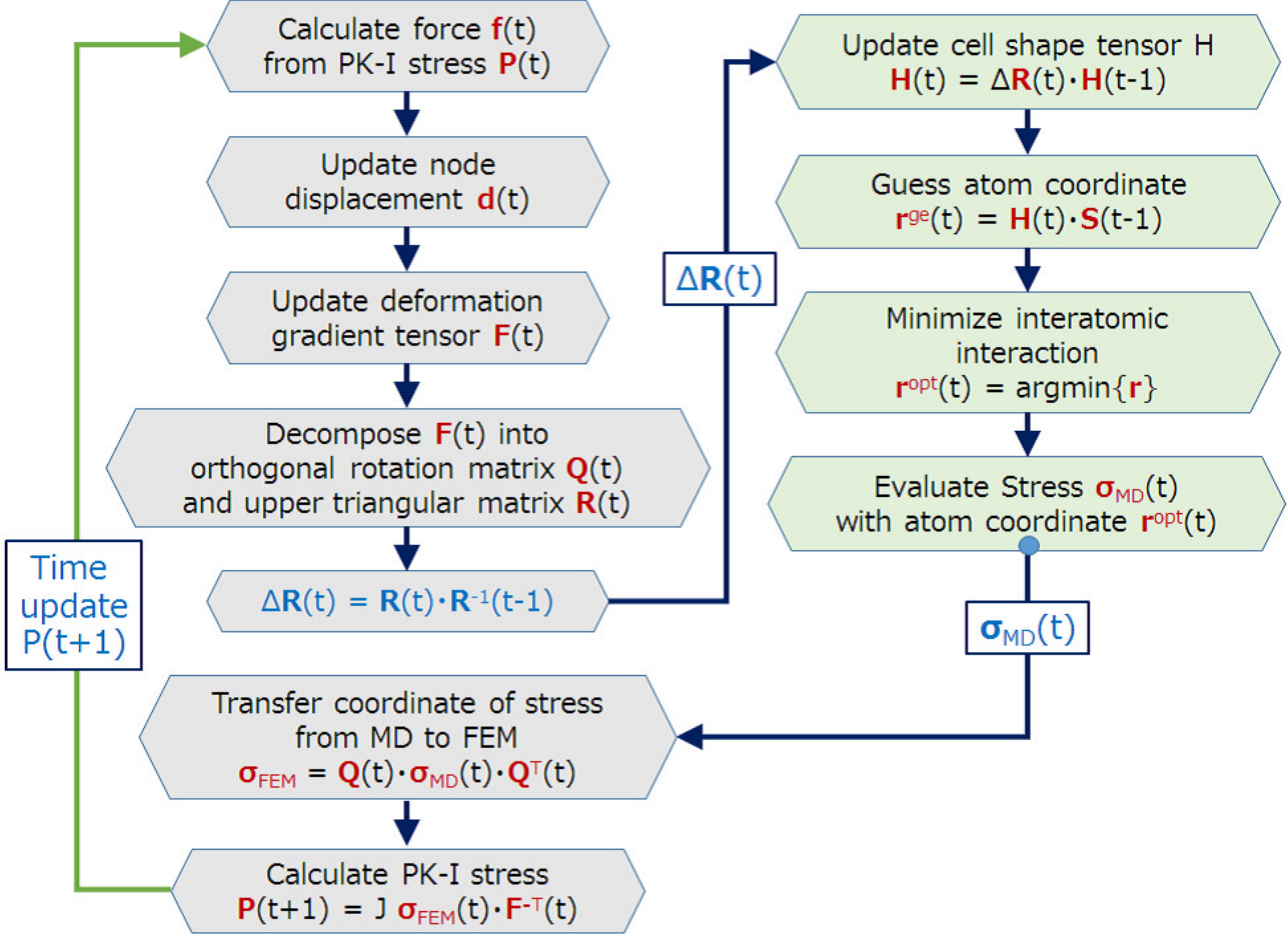}
  }
\end{center}
\caption{Flowchart of the multiscale computation algorithm that
  couples a continuum finite element simulation with a general MD simulation code.}
\label{fig:1}
\end{figure*}

The flow chart of the
proposed multiscale computation algorithm is schematically depicted in Fig. \ref{fig:1},
and its validity will be examined in Sec. \ref{sec:2.1}.

\subsection{Hierarchical coupling method for amorphous polymers}
\label{sec:1.2}

Here we briefly introduce the algorithm of CG-PR method, which has been developed
as a hierarchical multiscale modeling tool
for amorphous solids (see \cite{urata2018,urata2017b}).

In the framework of finite element method,
the discrete equations of motion for nodal displacements can be evaluated form the following equation,
\begin{equation}
\mathbf{M} \cdot \ddot{\mathbf{d}} + \mathbf{f}^{\rm int}(\mathbf{d})
=  \mathbf{f}^{\rm ext},
\end{equation}
where $\mathbf{d}$ is node displacement vector, and
$\mathbf{M}$, $\mathbf{f}^{\rm int}$, and $\mathbf{f}^{\rm ext} $
are the mass matrix, force vectors, and external force, respectively.
These quantities are defined as follows,
\begin{eqnarray}
& &\mathbf{M} = \sss_{e=1}^{n^e} \int_{\Omega_e} \rho_0 \mathbf{N}_e^{T} \cdot \mathbf{N}_{e} dV, \\
& & \mathbf{f}^{\rm int} = \sss_{e=1}^{n^e} \int_{\Omega_e} \mathbf{B}_e^{T} \cdot \mathbf{P}_{e}({\bf d}) dV, \label{eq.f} \\
& &\mathbf{f}^{\rm ext} = \sss_{e=1}^{n^e} \Bigr\{ \int_{\Omega_e} \mathbf{N}_e^{T} \cdot \mathbf{B}_{e} dV
+ \int_{\partial \Gamma_t} \mathbf{N}_e^{T} \cdot \bar{\mathbf{T}}_{e} dS \Bigl\},
\end{eqnarray}
where, $\sss$ is the element assemble operator acting over all elements;
$\rho_0$ is the material density; $\mathbf{N}_e$ is the element shape function matrix;
$\bar{\mathbf{T}}_{e}$ is the traction vector on the surface,
and $\mathbf{B}_e$ is the element strain-displacement matrix as,
\begin{equation}
\mathbf{B}_e := \Biggl[ \frac{\partial \mathbf{N}_e}{\partial X}  \Biggr]~.
\end{equation}

In Eq. (\ref{eq.f}), $\mathbf{P}_e$ is the first Piola-Kirchhoff stress,
which can be evaluated using interatomic interaction among atoms in an r-cell as follows,
\begin{eqnarray}
{\bf P} = {1 \over 2 \Omega_0} {\partial W \over
\partial {\bf F}}
= {1 \over 2 \Omega_0} \sum_{i, j} {\partial V (r_{ij}) \over \partial r_{ij} }
{ {\bf r}_{ij} \otimes {\bf R}_{ij}
\over r_{ij}}~,
\label{eq:PK1}
\end{eqnarray}
where ${\bf R}_{ij} = {\bf H} (0) \cdot {\bf S}_{ij}$ 
is the relative position vector between initial positions of atoms $i$ and $j$,
and ${\bf r}_{ij} = {\bf r}_j - {\bf r}_i$ is the position 
vector between current positions of atoms $i$ and $j$.
$V$ is interatomic potential as a function of distance $r_{ij}$ and
$\Omega_0$ is the volume of r-cell.
In the case of single crystal, atom position ${\bf r}$ is unique, but
this is not the case for amorphous solids.
Optimizing the potential energy respect to distance $r$,
we can obtain the optimized position vector ${\bf r}_i^{\rm opt}$ as,
\begin{eqnarray}
 {\bf r}_{i}^{\rm opt}(t) = {\rm argmin} {V({\bf r}_i^{\rm ge})}~,
  \label{eq:CGPR1}
\end{eqnarray}
where ${\rm argmin}$ stands for argument of the minimum,
${\bf r}^{\rm ge}_i (t)$ is the position guessed from previous configuration at time $t -1$,
\begin{eqnarray}
 {\bf r}^{\rm ge}_i (t)  &=& {\bf F}(t)_e \cdot {\bf H(0)}_e \cdot {\bf S}_i (t-1).
\end{eqnarray}
This equation is essentially based on the Cauchy-Born rule, because the guessed atom coordinates
are uniformly determined following the deformation gradient in an element.


Once we have found ${\bf r}_i^{\rm opt}$,
the following coordinates can be obtained.
\begin{eqnarray}
{\bf S}^{\rm opt}_i  &=& ({\bf F}(t)_e \cdot {\bf H(0)}_e)^{-1} \cdot {\bf r}^{\rm opt}_i, \\
{\bf R}^{\rm opt}_i &=& {\bf H(0)}_e \cdot {\bf S}_i^{\rm opt}.
\label{eq:CGPR3}
\end{eqnarray}
Then, the correct 1st Piola-Kirchhoff stress is eventually defined as follows,
\begin{eqnarray}
{\bf P} = {1 \over 2 \Omega_0} {\partial W \over
\partial {\bf F}}
= {1 \over 2 \Omega_0} \sum_{i, j} {\partial V (r^{\rm opt}_{ij}) \over \partial r^{\rm opt}_{ij} }
{ {\bf r}^{\rm opt}_{ij} \otimes {\bf R}^{\rm opt}_{ij}
\over r^{\rm opt}_{ij}}~.
\label{eq:CGPR2}
\end{eqnarray}

\begin{figure}
\begin{center}
\resizebox{0.40\textwidth}{!}{
  \includegraphics{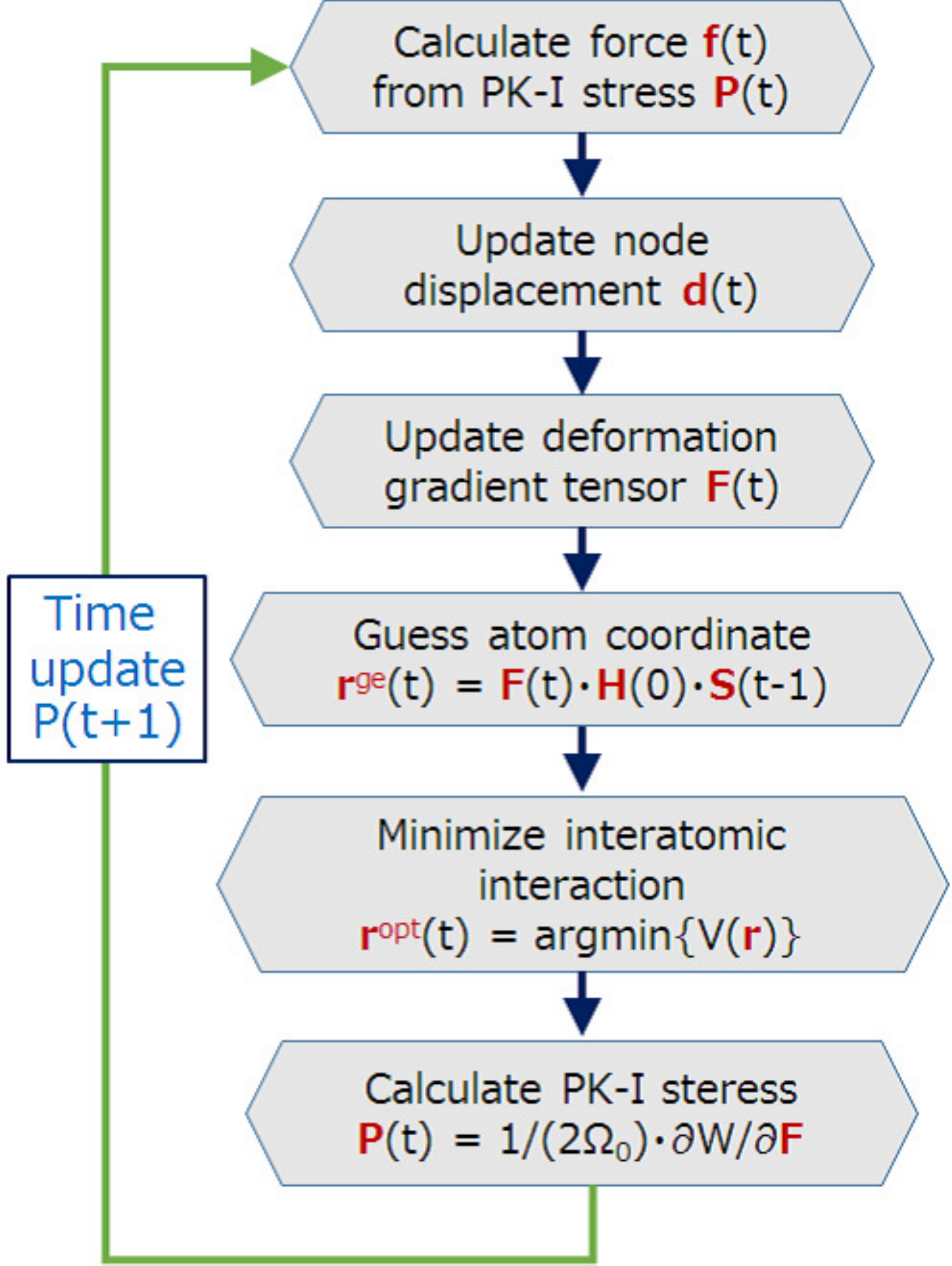}
}
\caption{Flowchart of the Coarse-grained Parrinello-Rahman (CG-PR) method.
\cite{urata2018,urata2017b}}
\label{fig:2}
\end{center}
\end{figure}

If we use a general MD code, the 1st Piola-Kirchhoff stress is calculated using
Eq.(\ref{eq:Pmd}) with the optimized coordinate ${\bf r}^{\rm opt}$.
The flowchart of the above algorithm is shown in Fig. \ref{fig:2}.

\subsection{Implementation of parallel MD systems in LAMMPS}
\modifym{}{
In the proposed multiscale model,
each element at continuum FEM scale
has its own MD system, i.e. fine scale r-cells
at every FEM quadrature points.
LAMMPS can divide the MD system into multiple systems so that these systems
are independent from others
by splitting a message passing interface (MPI) communicator
to multiple communicators assigned to each MD systems.
}
These MD systems are independent at the MD simulation level,
while exchanging the energy and momentum among them through FEM.
Since the size of FEM is small in the present work,
FEM is computed on a parent node, and LAMMPS is computed on full nodes.
To communicate FEM and LAMMPS, simple scatter and gather algorithm has been implemented.
The cost of the MPI communication is very small, and we shall discussed it in Sec. \ref{sec:2.2}.

\section{Results}
\label{sec:2}

\subsection{Model validation}
\label{sec:2.1}

\begin{figure}
\begin{center}
\resizebox{0.48\textwidth}{!}{
  \includegraphics{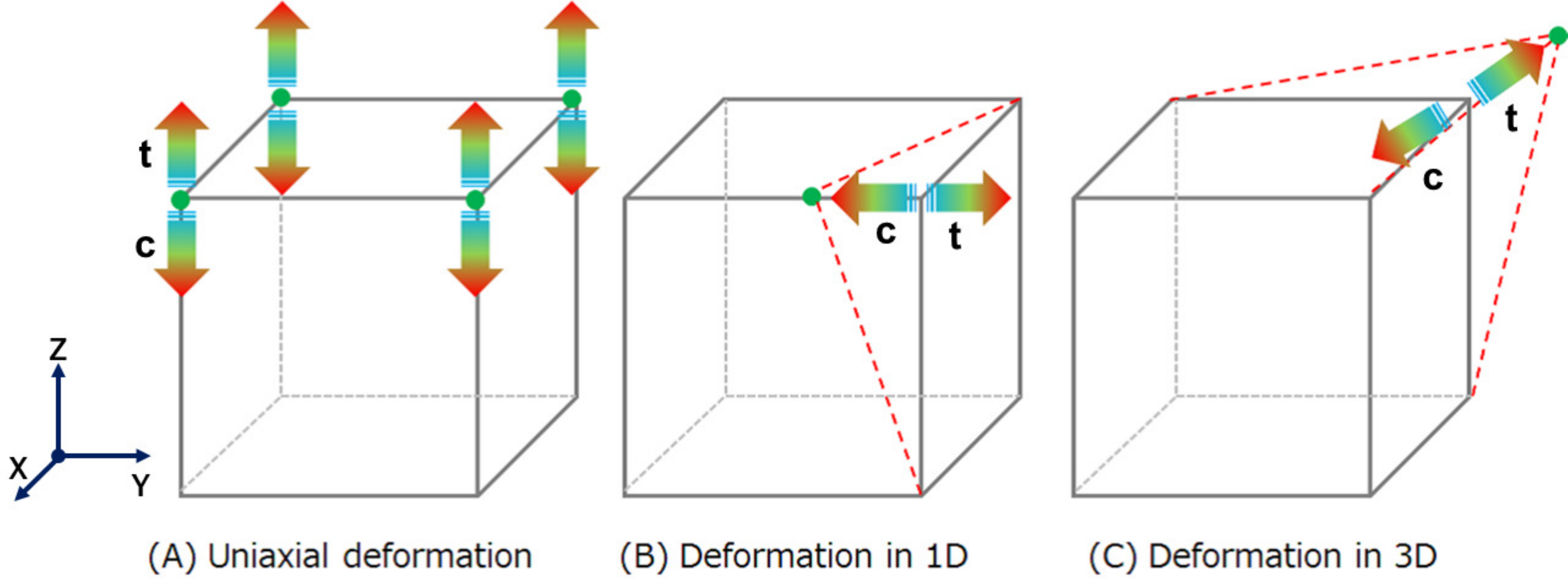}
}
\caption{Deformation patterns used for validation tests. \modifym{}{``t'' and ``c'' represent tensile and compress.}}
\label{fig:3}
\end{center}
\end{figure}

\begin{figure}
\begin{center}
\resizebox{0.4\textwidth}{!}{
  \includegraphics{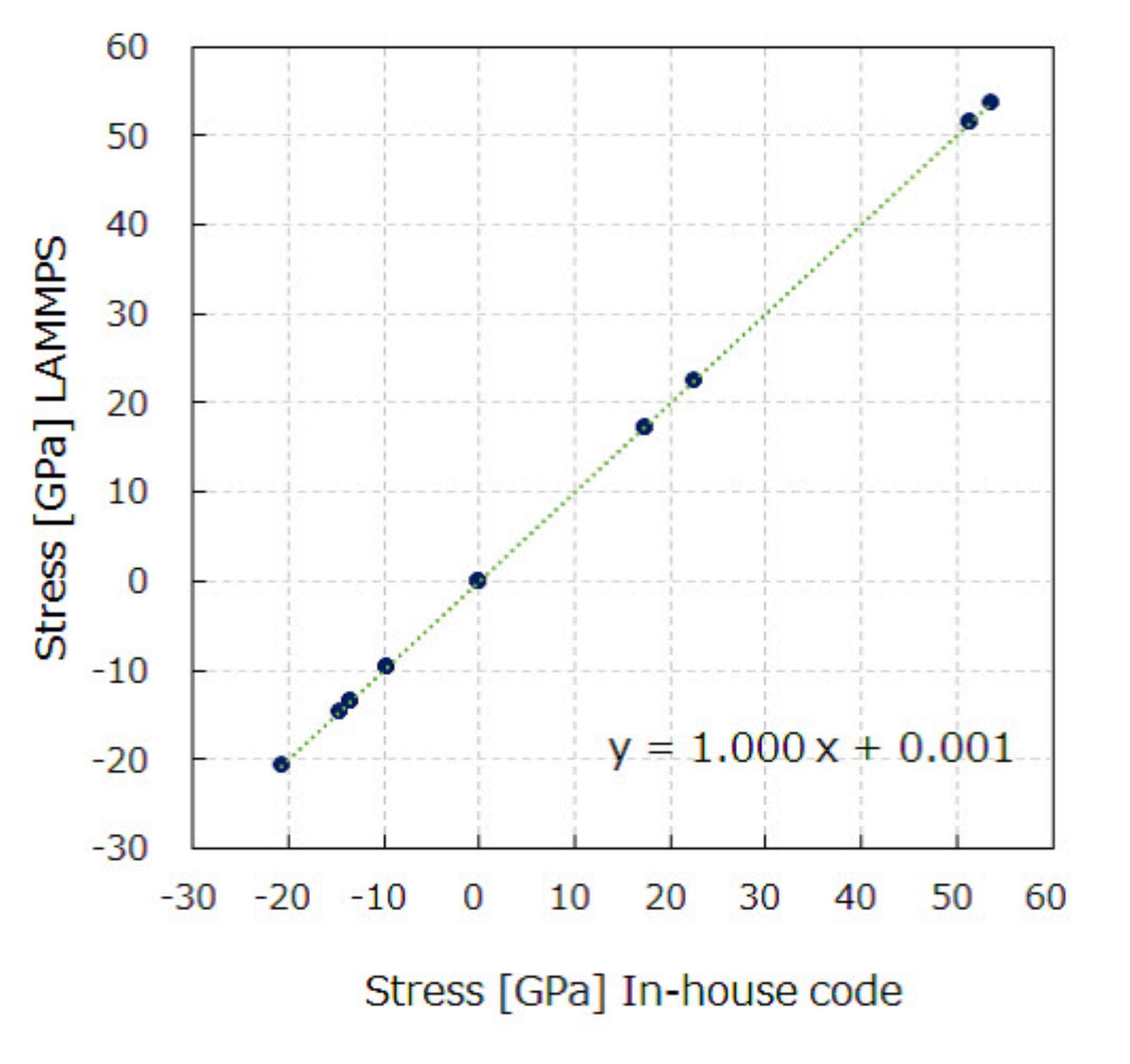}
}
\caption{Comparison of stresses evaluated using two algorithms for simple deformation (Fig. \ref{fig:3} (A)).
One is coupling method with a general MD code using Eqs.(\ref{eq:LAMMPS1})--(\ref{eq:LAMMPS2}),
and the other one is original in-house code with Eqs.(\ref{eq:CGPR1})--(\ref{eq:CGPR2}).
}
\label{fig:4}
\end{center}
\end{figure}

First of all, to test our multiscale code that couples the open-source MD code LAMMPS with FEM method,
we compare the stresses evaluated by the two independent algorithms mentioned above,
i.e. the one that uses Eqs.(\ref{eq:CGPR1})--(\ref{eq:CGPR2}) and
the other one that uses LAMMPS with Eqs.(\ref{eq:LAMMPS1})--(\ref{eq:LAMMPS2}).
A crystalline copper model, which is composed of $10 \times 10 \times 10$  FCC primitive cells,
is used for the validation test.
The supercell includes 4,000 atoms and the side length is 36.15 \AA.
\modifym{Mashin's}{Mishin's} EAM potential \cite{mishin2001} was employed to evaluate interatomic interaction.

Uniaxial stretching/compression and two kinds of asymmetric deformations
were examined as schematically illustrated in Fig. \ref{fig:3}.
In the first case, the unit cell was simply elongated and compressed along z-axis
with maintaining the cell length of the other two sides.
The stresses in every components evaluated using two methods are
compared in a scatter diagram in Fig. \ref{fig:4},
confirms that the stresses calculated by the two codes agree well
for the symmetric deformation.

Next, we examined how the multiscale algorithm works when \modifym{}{it is}
applied to simulate asymmetric deformations,
in order to verify the multiscale algorithm ability to
use QR decomposition to couple with a MD engine,
which is LAMMPS in this case.
In Case (B), one edge is displaced from -0.3 to +0.3 strain along $y$-axis,
whereas the edge is moved within the same range of strain
for \modifym{all directions}{the [111]-direction} at the same time \modifym{}{in the case of Case (C).}
In Table \ref{tab:1}, we summarize the deformation gradient tensor ${\bf F}$,
the orthogonal rotation tensor ${\bf Q}$, and the upper triangular stretch tensor ${\bf R}$
in addition to all stress components in both coordinate systems of LAMMPS and FEM.
One may find that all stresses calculated by using QR decomposition
(Eqs.(\ref{eq:LAMMPS1})--(\ref{eq:LAMMPS2}).), $\bsigma_{\rm FEM}$,
agree to those of the in-house CG-PR code with Eqs.(\ref{eq:CGPR1})--(\ref{eq:CGPR2}),
$\bsigma_{\rm IH}$. It is therefore reasonable to expect that
the proposed multiscale coupling algorithm
may be able to simulate even more complicated asymmetric deformations,
which may not be feasible to simulate
by using the Parrinello-Rahman method \cite{parrinello1980}
with the periodic boundary conditions.

In the next section, we shall demonstrate a large scale multiscale simulation
of inelastic deformation of a polymeric material.

\begin{table*}
\caption{Comparison of stresses  [GPa] evaluated using in-house code $\bsigma_{\rm IH} $ and
the coupling method with LAMMPS $\bsigma_{\rm FEM}$ for deformation drawn
in Fig.~\ref{fig:3} (B) \modifym{}{and (C)}.
\modifym{}{In Case (B), one edge is displaced with 0.3 or -0.3 strain along $Y$-axis.
In Case (C), one edge is moved in 0.3 or -0.3 strain along the [111]-direction.
(t) and (c) stand tensile and compressive loadings, respectively.}
Deformation gradient tensor {\bf F} and the decomposition to
the orthogonal rotation matrix ${\bf Q}$
and the upper triangular stretch matrix ${\bf R}$ are also displayed.
\modifym{}{
The last line shows the $L2$-norm (spectral norm) of the error between
$\boldsymbol{\sigma}_{\rm IH}$ and $\boldsymbol{\sigma}_{\rm FEM}$.}
 }
\label{tab:1}
\begin{tabular}{c | c c c | c c c | c c c | c c c }
\hline\noalign{\smallskip}
Fig.~\ref{fig:3}    &    \multicolumn{3}{c}{B(t)} &   \multicolumn{3}{c}{B(c)} &  \multicolumn{3}{c}{C(t)} &   \multicolumn{3}{c}{C(c)}  \\
\noalign{\smallskip}\hline\noalign{\smallskip}
        &  1.000 & 0.000 & 0.000 &  1.000 & 0.000 &  0.000 & 1.075 &  0.075 & 0.075 & 0.925 & -0.075 & -0.075 \\
{\bf F} & -0.075 & 1.075 & 0.075 &  0.075 & 0.925 & -0.075 & 0.075 & 1.075 & 0.075  & -0.075 & 0.925 & -0.075 \\
        &  0.000 & 0.000 & 1.000 &  0.000 & 0.000 &  1.000 & 0.075 & 0.075 & 1.075  & -0.075 & -0.075 & 0.925 \\
\noalign{\smallskip}\hline\noalign{\smallskip}
        &  0.997 & 0.075 & 0.000 & 0.997 & -0.075 & 0.000 & 0.995 & -0.074 & -0.065 & 0.993 &  0.073 & 0.088 \\
{\bf Q} & -0.075 & 0.997 & 0.000 & 0.075 &  0.997 & 0.000 & 0.069 & 0.995 & -0.065   & -0.081 & 0.993 & 0.088 \\
        &  0.000 & 0.000 & 1.000 & 0.000 &  0.000 & 1.000 & 0.069 & 0.060 & 0.996  & -0.081 &  -0.094 & 0.992 \\
\noalign{\smallskip}\hline\noalign{\smallskip}
        & 1.003 & -0.080 & -0.006 & 1.003 & 0.069 & -0.006 & 1.080 & 0.154 & 0.154 & 0.931 & -0.143 & -0.143 \\
{\bf R} & 0.000 &  1.072 &  0.075 & 0.000 & 0.922 & -0.075 & 0.000 & 1.069 & 0.134   & 0.000 & 0.920 & -0.167 \\
        & 0.000 &  0.000 &  1.000 & 0.000 & 0.000 &  1.000 & 0.000 & 0.000 & 1.061  & 0.000 & 0.000 & 0.905 \\
\noalign{\smallskip}\hline\noalign{\smallskip}
                             & -7.536 &  3.844 & 0.731 & 12.786 & -8.959 & 1.312 & -15.403 &  -5.232 & -4.614 & 75.792 & 17.495 & 21.018 \\
$\bsigma_{\rm MD}$  & 3.844 & -9.429 & -3.615   & -8.959 & 18.757 & 9.232 & -5.232 & -13.891 & -3.995   & 17.495 & 81.230 & 24.536 \\
                             & 0.731 & -3.615 & -6.974 & 1.312 & 9.232 & 14.155 & -4.614 & -3.995 & -12.884 & 21.018 & 24.536 & 90.283 \\
\noalign{\smallskip}\hline\noalign{\smallskip}
                             & -6.974 &  3.660 & 0.458 & 14.155 & -9.304 & 0.618  & -14.060 &  -4.698  & -4.698 & 82.435 & 21.629 & 21.629 \\
$\bsigma_{\rm FEM}  $       & 3.660 & -9.992 & -3.660   & -9.304 & 17.387 & 9.304 & -4.698 & -14.060 & -4.698  & 21.629 & 82.435 & 21.629 \\
                             & 0.458 & -3.660 & -6.974 & 0.618 & 9.304 & 14.155 &  -4.698 & -4.698 & -14.060 & 21.629 & 21.629& 82.435 \\
\noalign{\smallskip}\hline\noalign{\smallskip}
\noalign{\smallskip}\hline\noalign{\smallskip}
                             & -6.975 &  3.660 & 0.458 & 14.154 & -9.304 & 0.618 & -14.061 &  -4.698 & -4.698 & 82.433 & 21.629 & 21.629 \\
$\bsigma_{\rm IH}  $       & 3.660 & -9.993 & -3.660   & -9.304 & 17.386 & 9.304 &  -4.698 & -14.061 & -4.698   & 21.629 & 82.433 & 21.629 \\
                          & 0.458 & -3.660 & -6.975 & 0.618 & 9.304 & 14.154 &  -4.698 & -4.698 & -14.061 & 21.629 & 21.629 & 82.433 \\
\noalign{\smallskip}\hline\noalign{\smallskip}
\noalign{\smallskip}\hline\noalign{\smallskip}
$||\Delta \boldsymbol{\sigma}||_2$ & &0.001& & &0.001& & &0.001& & &0.002& \\
\noalign{\smallskip}\hline\noalign{\smallskip}
\end{tabular}
\end{table*}

\subsection{Large-scale parallel multiscale computation for polymer deformation}
\label{sec:2.2}

\begin{figure}
 \begin{center}
 \resizebox{0.45\textwidth}{!}{
   \includegraphics{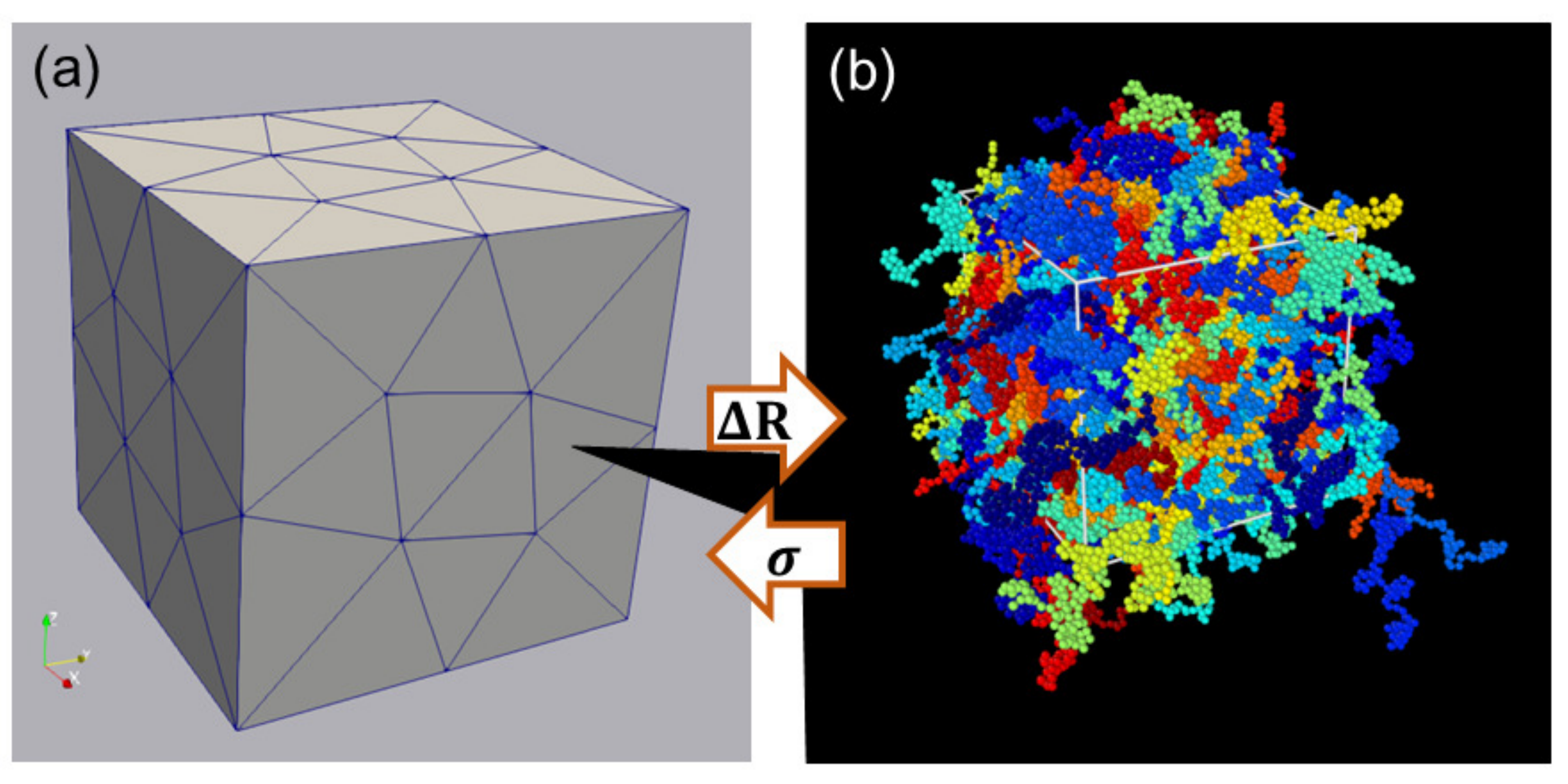}
 }
 \caption{Schematic illustration of multiscale simulation for a polymeric material:
 a cubic material with finite element mesh (a) and polymers in a r-cell (b).}
 \label{fig:5}
 \end{center}
 \end{figure}

\begin{figure*}
 \begin{center}
 \resizebox{\textwidth}{!}{
   \includegraphics{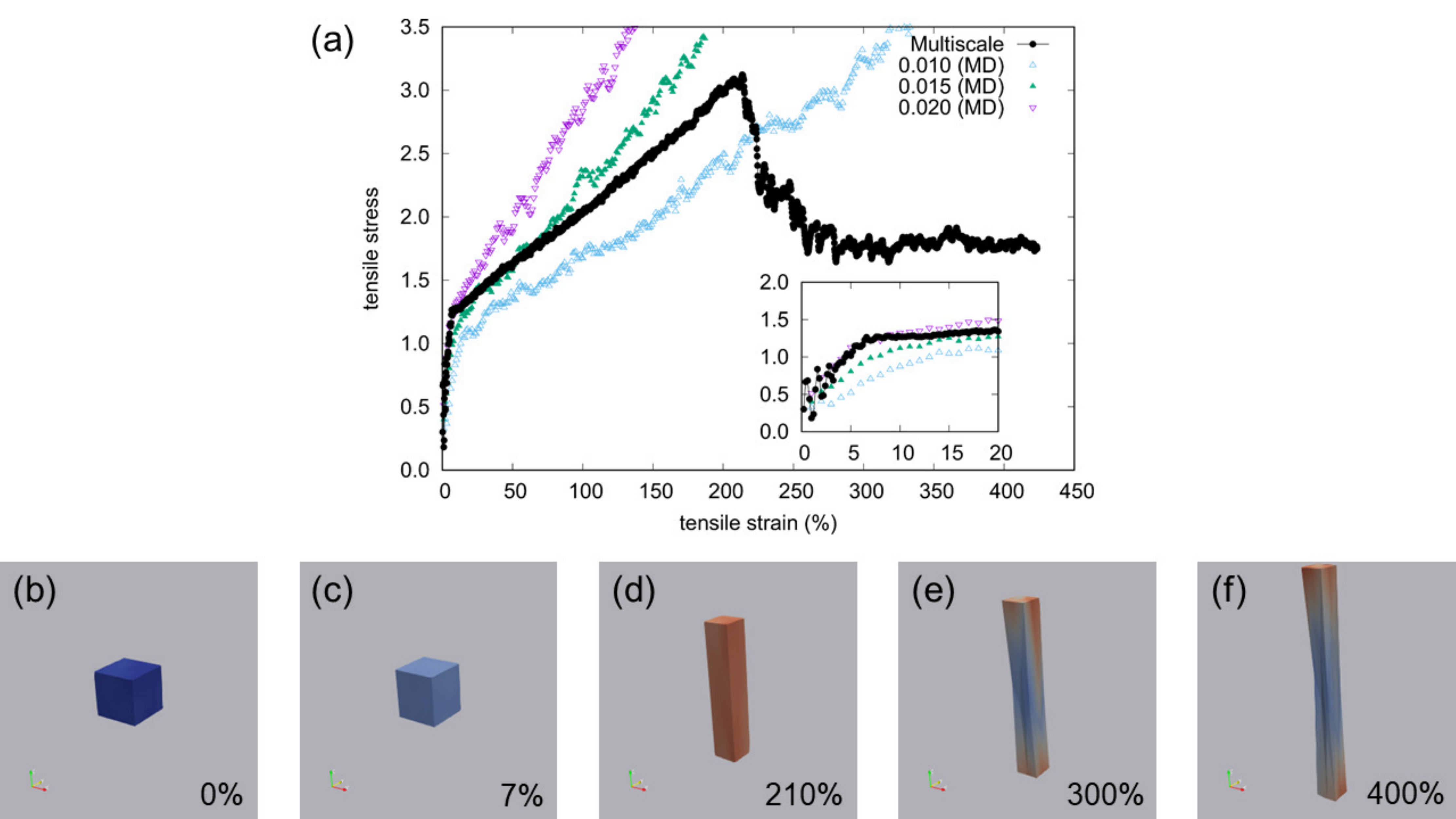}
 }
 \caption{The tensile stress versus the tensile \modifym{ratio}{strain} of uniaxially stretched polymeric material \modifym{}{obtained from the multiscale simulation} (a).
   The unit of the stress here is $[\epsilon/\sigma^3]$, $\epsilon$ and $\sigma$ are the unit of energy and the unit of size of LJ particle in a coarse-grained polymer, respectively.
   \modifym{}{Molecular dynamics simulation results under different uniaxial engineering strain rates (0.01, 0.015, 0.020 $[1/\tau]$) are also shown in Fig (a). Inset in Fig (a) focuses on the transition from elastic to plastic deformation.}
Figures (b) to (f) represent the appearances of the material during deformation at characteristic \modifym{points shown}{strain found} in Fig. (a). The color \modifym{}{shown in Figs. (b) to (f)} represents the magnitude of the stress at the local finite element; blue is low and red is high. }
 \label{fig:6}
 \end{center}
 \end{figure*}

In this section,
we report the results of using the coupled FEM-LAMMPS code to investigate inelastic
deformation of a macroscale cubic polymeric specimen composed of many mono-disperse linear polymers.
A schematic illustration of the numerical model is shown in Fig. \ref{fig:5}.
The coarse-grained polymer model, namely, the bead-spring model~\cite{KremerGrest1990},
was used to describe the polymer chains in r-cells.
The polymer chain is composed of LJ particles linearly connected
by finite extensively nonlinear elastic bonds.
The potential parameters were set to the standard Kremer-Grest model~\cite{KremerGrest1990}.
The cubic specimen is composed of $100$ tetrahedron finite elements, and each element has
an r-cell that contains $100$ polymer chains, and each of them is composed of $100$ beads.
In FEM calculation, each quadrature point is represented by an r-cell.
Because of the linear four-node tetrahedron element is used in FEM modeling,
each element has only one quadrature point. Thus,
the total number of beads in fine scale model of the system is $100 \times 100 \times 100 =10^6$,
i.e. one million atomistic freedoms.
Initial states of polymers were prepared by using OCTA-COGNAC~\cite{OCTA-COGNAC},
where the polymers were equilibrated in melt condition.
To impose the uniaxial stretch boundary condition to the polymeric material,
the constant velocities were prescribed at both top and bottom surfaces
of the cubic specimen along vertical direction.
The sides of the material were free to move.
The magnitude of the prescribed velocity was chosen sufficiently large
so that one can observe nonlinear behaviors of the polymeric material.
When the deformation rate is higher than a characteristic value $1/\tau_{\rm R}$,
where $\tau_{\rm R}$ is the Rouse relaxation time of polymer chains,
the polymeric material is expected to show a nonlinear behavior in an extensional deformation~\cite{Murashima2018}.
In the case of polymer melt with 100 beads per a chain, $\tau_{\rm R} \simeq 10,000 [\tau]$,
where $\tau$ is the unit of time of LJ particle.
The results are summarized in Fig. \ref{fig:6}.

 Figure \ref{fig:6}(a) shows the tensile stress versus the tensile \modifym{ratio}{strain},
 \modifym{
where the tensile stress and the tensile ratio defined here
are the magnitude of the largest eigenvalue of the stress tensor $\bsigma_{\rm FEM}$
and the deformation gradient tensor ${\bf F}$, respectively, averaged over the whole system.}
        {where the tensile stress defined here is the magnitude of the largest eigenvalue of the stress tensor $\bsigma_{\rm FEM}$ averaged over the whole system. The tensile strain was estimated from the uniaxial engineering strain rate $2.0\times 10^{-5}$ [1/FEM step] multiplied by the FEM steps.}
The unit of the stress is [$\epsilon/\sigma^3$], where $\epsilon$ and $\sigma$
are the units of the energy and the length
of LJ coarse-grained polymer chain, respectively.
\modifym{}{For comparison, we carried out the molecular dynamics simulation of the Kremer-Grest model on the periodic boundary conditions, namely the unit cell of the multiscale simulation. We applied uniaxial engineering strain rates ($0.01, 0.015, 0.02 [1/\tau]$) to the system at $T=0.1$ which was selected since the temperature dependence of the stress-strain curve is negligible when the temperature is less than the glass transition temperature 0.35~\cite{Bennenman1998}. The results of the molecular dynamics simulation were also shown in Fig. \ref{fig:6}(a).
}
Figures \ref{fig:6}(b) to (f) show the appearances of the material
at \modifym{the characteristic tensile ratios, 1.0 (initial state) (b), 1.08 (c), 3.14 (d), 4.0 (e), 5.0 (f).}{the characteristic tensile strain; 0\% (initial state, b), 7\% (first yield point, c), 210\% (second yield point, d), 300\% (lower yield point, e), 400\% (final state, f).
The double yield points observed in polymeric materials have been investigated and the second yield point relates to the necking phenomena~\cite{Brooks1992}.}
The tensile stress increases monotonically from (b) to (c) and from (c) to (d).
\modifym{}{
  The first yield point (c) is the onset of the plastic deformation.
  }
The former region is the elastic deformation region and the latter region is the plastic deformation region.
The shape of the material is a rectangular parallelepiped in these small strain regions.
Beyond the \modifym{}{second} yield point (d), the tensile stress decreases and the material show necking.
\modifym{}{
The first yield point appeared around
the 7\% strain
both in the multiscale simulation and the molecular dynamics simulation 
with the periodic boundary conditions.
When the strain is higher than the first yield point, 
the stress-strain curves show different slopes owing to its strain-rate dependence.
The molecular dynamics simulation did not show the second yield point that is around the strain 200\%.
The reason for the difference between the multiscale simulation and the molecular dynamics simulation 
is that the molecular dynamics simulation was carried out under periodic boundary condition 
without the free-surfaces, which prevents necking.
However, even if the large scale molecular dynamics simulation with the surfaces were carried out, 
the necking behavior was not observed in the amorphous polymeric material~\cite{Yashiro2003}.}
\modifym{}{
On the other hand,
the coarse-grained molecular dynamics simulation of polyethylene lamellar structures, 
consisting of crystalline and amorphous layers, has shown a necking like behavior,
where a neck is formed in the weak amorphous layers and propagates down in the specimen
~\cite{Higuchi2017}.
It is believed that
inhomogeneity may play a key role in formation of necking in amorphous polymeric materials.
Capturing necking formation in the amorphous polymer in the molecular dynamics simulation 
is still a challenging problem.
The present multiscale solution, however, 
has somewhat overcome this difficulty in the large scale hierarchical
multiscale simulation.
}

The material behaviors observed from Fig. \ref{fig:6} 
are qualitatively consistent with the experimental observation 
from an uniaxially melt-extruded high-density polyethylene films,
e.g.~\cite{ZhouWilkes1998}.
The \modifym{}{second} yield point has been observed when \modifym{the tensile ratio 
is about three, namely the engineering strain is 200\%,
}{the engineering strain is about 200\%} both in our simulation 
and in the experiment reported.
Nevertheless, 
the initial condition of the polymer specimen in the simulation was 
in amorphous state while that in the experiment was crystalline lamellae.
It was also speculated that 
the yield point may depend on the sample size and the direction of stretch
(see \cite{ZhouWilkes1998}).
Further investigation, adjusting the conditions between the simulation and the experiment, is still ongoing.

\begin{figure}
 \begin{center}
 \resizebox{0.4\textwidth}{!}{
   \includegraphics{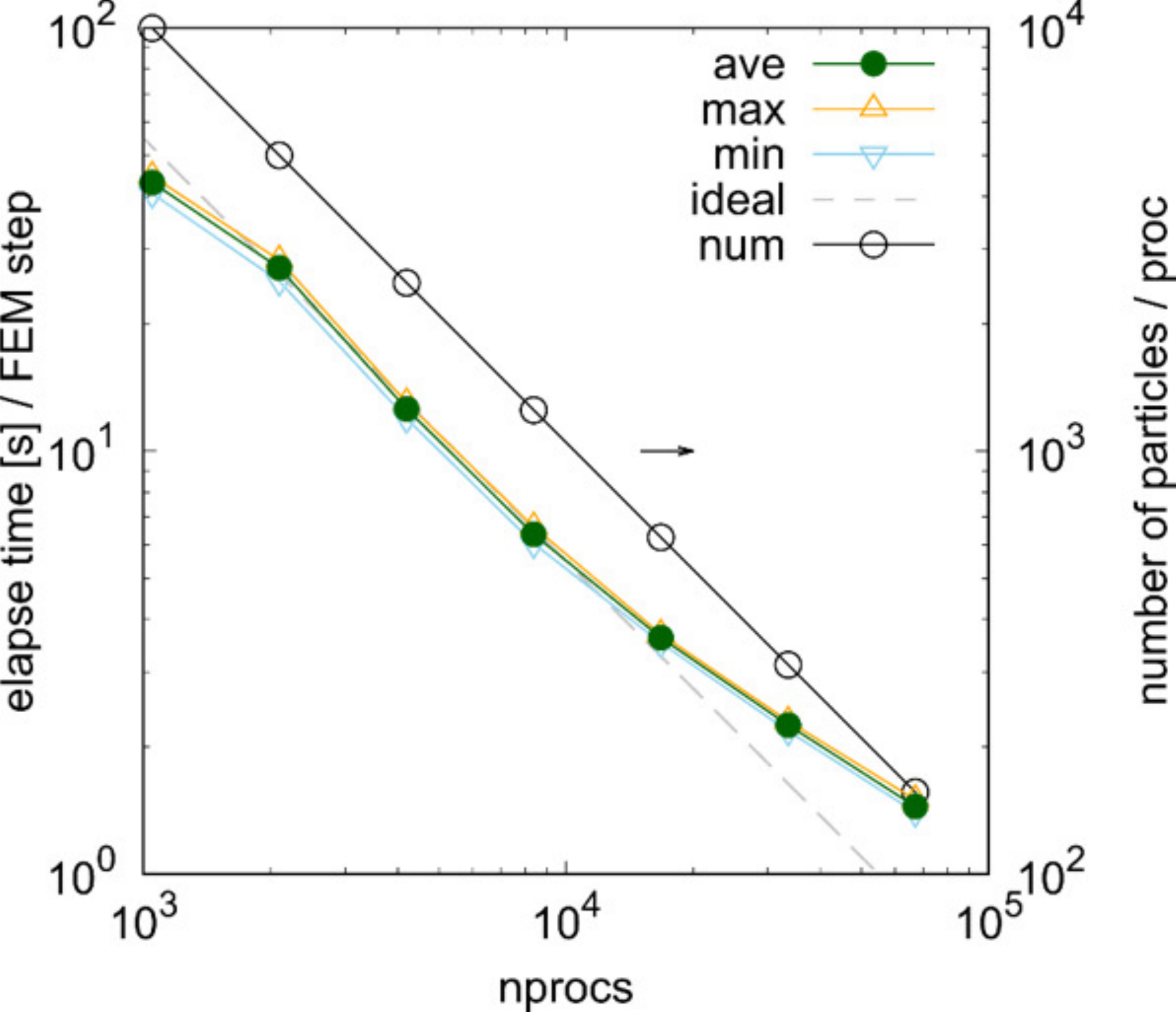}
 }
 \caption{Parallel scaling of multiscale simulation on K computer. The simulation system was composed of one thousand finite elements and each finite element had 10 thousand LJ particles (100 bead-spring chains with 100 beads). \modifym{}{The items, ``ave'', ``max'' and ``min'', are the average, maximum, and minimum times (left vertical axis) in 10 multiscale FEM steps, respectively, ``ideal'' is the ideal parallel scaling line for guide to eyes, and ``num'' is the number of particles per a processor (right vertical axis). Note that each MD simulation in the multiscale simulation is not parallelized when the number of processors is $10^3$.}}
 \label{fig:7}
 \end{center}
 \end{figure}

\modifym{}{
Moreover, we have studied 
parallel scalability of our multiscale code on K computer~\cite{Knotes}.
In the study, the total number of elements in the FEM model
is up to one thousand, 
and the total number of particles in a system is up to 10 millions.}
Elapsed CPU time [unit s] per a multiscale FEM time step was measured and averaged over 10 FEM steps,
which includes the iteration time for running the fine scale CG-PR method in each r-cell.
As maybe found in Fig. \ref{fig:7}, our multiscale
simulation result 
shows a nearly ideal scaling efficiency 
when the number of particles per a processor is larger than 100.
Generally speaking, parallel scalability of a direct MD simulation 
with 10 million particles saturates when the number of processors reaches to 10 thousands.
The reason for the
high scaling efficiency of the hybrid method implemented with FEM and LAMMPS may be attributed to
the facts that (1) the multiscale method, especially the natural segregated distribution of r-cells,
makes efficient use of the massively parallel architecture of the supercomputer used
such as K computer, and (2) the CG-PR molecular statics minimization algorithm may have less or no
computing overhead.

\section{Conclusions}
\label{sec:3}
In this work, we have developed a hierarchical multiscale
method to bridge finite element method with molecular dynamics
by using the large scale atomic /molecular massively parallel simulator (LAMMPS).
In the developed computation algorithm, 
a QR decomposition is used and implemented to couple deformations with stresses at both
FEM level as well as MD level, and in between.
The multiscale coupling algorithm was validated by applying typical deformation patterns 
on Copper crystal model, and a good agreement between MD simulation results 
with the results obtained from the CG-PR method is confirmed.

\modifym{}{
We have applied the multiscale code simulating uniaxial deformation of
a polymeric material in a multiscale CG-PR/FEM model that is composed
of one million fine scale r-cell of Lennard-Jones particles,
and we have succeeded in observing nonlinear elastic-plastic deformation behavior of
amorphous polymeric materials,
including strain localization (almost necking) in the polymeric material. 
}
Indeed, a yield point has appeared
when \modifym{the tensile ratio is about three}{the tensile strain is about 200\%}, 
which is consistent with the experimental observation.

It is worthy noting that the four-node tetrahedral element (C3D4)
has a relative lower interpolation
order, and it is often too stiff to be used in inelastic finite deformation
computations. The inelastic deformation simulation results reported
in this work by using C3D4 element is remarkable,
which could not be achieved by using the same type of element in
the phenomenological computational plasticity.
\modifym{}{In the proposed multiscale computational algorithm, 
the r-cell at each Gauss point 
can change the volume
in accordance with the stress field through the hierarchical coupling. 
The comparability of the r-cell may avoid the locking issue caused in the usual FEM model.}

In addition, we also examined parallel scalability of the multiscale code
using a large FEM model, in which ten million fine scale particles are embedded.
The result
revealed that the code has a strong, nearly ideal, scaling ability
up to ten thousand processors thanks to the highly efficient built-in scalability of LAMMPS.
The large scale simulations reported in this paper
demonstrated applicability of the multiscale coupling method and
algorithm to model and simulate mechanical properties \modifym{}{of}
amorphous materials.

\section*{Authors' contributions}
TM designed the concurrent coupling algorithm and developed a code to establish
interface with LAMMPS.
SU developed the multiscale simulation code coupling FEM and LAMMPS,
and performed validation tests.
TM conducted large scale simulation for the polymeric material.
TM and SU wrote simulation parts of the manuscript,
and SL supervised the project, wrote and summarized the manuscript.
All the authors have read and approved the final manuscript.

\section*{Acknowledgment}
\begin{acknowledgement}
TM would like to thank Professor M. Isobe who brought TM and SU together.
Our collaboration would not have started without his help.
 TM also thanks to
 \modifym{}{Professor K. Yashiro who gave us several comments on the large scale molecular dynamics simulation,}
Professor  T. Kawakatsu and Professor T. Taniguchi for their supports and fruitful discussions.
TM was supported by MEXT as ``Exploratory Challenge on Post-K computer'' (Challenge of Basic Science - Exploring Extremes through Multi-Physics and Multi-Scale Simulations).
This research used the computational resources of the K computer provided by the RIKEN Advanced Institute for Computational Science through the HPCI System Research project (Project ID: hp160267, hp170236, hp180176, hp190186) and the facilities of the Supercomputer Center in the Institute for Solid State Physics at the University of Tokyo.
\end{acknowledgement}

%

 \bibliographystyle{epj}

%
%
%

\end{document}